\documentclass[useAMS,usenatbib]{mn2e}

\usepackage{epsfig}
\usepackage{graphicx}

\title[ISINA]
{ISINA: INTEGRAL Source Identification Network Algorithm\thanks{Based on observations with {\it INTEGRAL}, an ESA project with instruments and science data centre funded by ESA member states (especially the PI countries: Denmark, France, Germany, Italy, Spain), Czech Republic, and Poland, and the participation of Russia an the USA.}}
\author[S. Scaringi, A.J. Bird, D.J. Clark, A.J. Dean, A.B. Hill, V.A. McBride and S.E Shaw]
{S. Scaringi$^{1}$\thanks{E-mail: simo@astro.soton.ac.uk},
A.J.Bird$^{1}$, D.J. Clark$^{1}$, A.J. Dean$^{1}$, A.B. Hill$^{1}$, V.A. McBride$^{1}$ and S.E. Shaw$^{1}$\\ $^{1}$Department of
Physics and Astronomy, University of Southampton, Highfield, SO17 1BJ, UK}

\begin{document} 

\date{}

\pagerange{\pageref{firstpage}--\pageref{lastpage}} \pubyear{2008}

\maketitle

\label{firstpage}

\begin{abstract}
We give an overview of ISINA: INTEGRAL Source Identification Network Algorithm. This machine learning algorithm, using Random Forests, is applied to the IBIS/ISGRI dataset in order to ease the production of unbiased future soft gamma-ray source catalogues. First we introduce the dataset and the problems encountered when dealing with images obtained using the coded mask technique. The initial step of source candidate searching is introduced and an initial candidate list is created. A description of the feature extraction on the initial candidate list is then performed together with feature merging for these candidates. Three training and testing sets are created in order to deal with the diverse timescales encountered when dealing with the gamma-ray sky. Three independent Random Forest are built: one dealing with faint persistent source recognition, one dealing with strong persistent sources and a final one dealing with transients. For the latter, a new transient detection technique is introduced and described: the Transient Matrix. Finally the performance of the network is assessed and discussed using the testing set and some illustrative source examples.
\end{abstract}


\section{Introduction}

\begin{figure}
\centering
\includegraphics[width=0.5\textwidth]{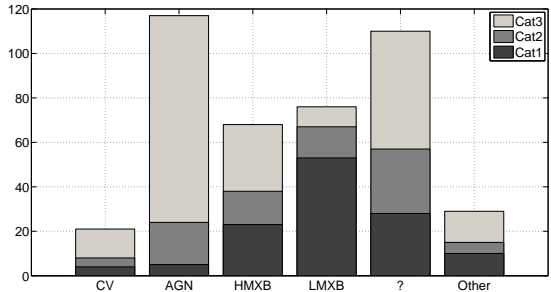}	
\caption{Numbers of sources in the first, second, and third IBIS/ISGRI catalogues, classified by type: Cataclysmic Variables (CV), Active Galactic Nuclei (AGN), High Mass X-ray Binaries (HMXB), Low Mass X-ray Binaries (LMXB), Other and Unknowns (?)}
\label{fig:source_dist}
\end{figure}

Since its launch in 2002, the INTEGRAL (International Gamma-Ray Astrophysics Laboratory) satellite has carried out more than 5 years of observations in the energy range from 5 keV to 10 MeV. In particular the IBIS (Imager on Board INTEGRAL spacecraft) imaging instrument (\citealt{ubertini03}) has been optimised for survey work, with a large ($30^o$) field of view with excellent imaging and spectroscopic capability, and has formed the basis of several previous INTEGRAL surveys. Because of the success of INTEGRAL, with longer exposure times and larger sky coverage, the amount of data accumulated over the mission lifetime has dramatically increased, and with it the number of detected objects (see Fig.\ref{fig:source_dist}). Moreover different types of objects are also being detected, with a high ratio of unknowns.  

Conventional methods for the production of source catalogues are becoming less and less adequate, requiring the use of novel techniques for the reduction of visual inspection required for the compilation of future catalogues. Moreover novel algorithms for classification and identification will produce more objective and homogeneous sets compared to the subjective samples obtained through visual inspection.  In this paper we will describe the implementation of a supervised machine learning algorithm which has been developed for the purpose of source identification on the IBIS/ISGRI images based on Random Forests (\citealt{randfo}). In machine learning, a Random Forest is a classifier that builds many single decision trees, and uses these to output a predicted class for a given input. Each single decision tree produces a predicted class for a given input, whilst the Random Forest determines the mode of all these small predictors to produce a more accurate classifier. The algorithm will be described in more detail, but in order to fully appreciate it the data being classified must be thoroughly understood.

\begin{figure*}
\centering
\includegraphics[width=0.75\textwidth]{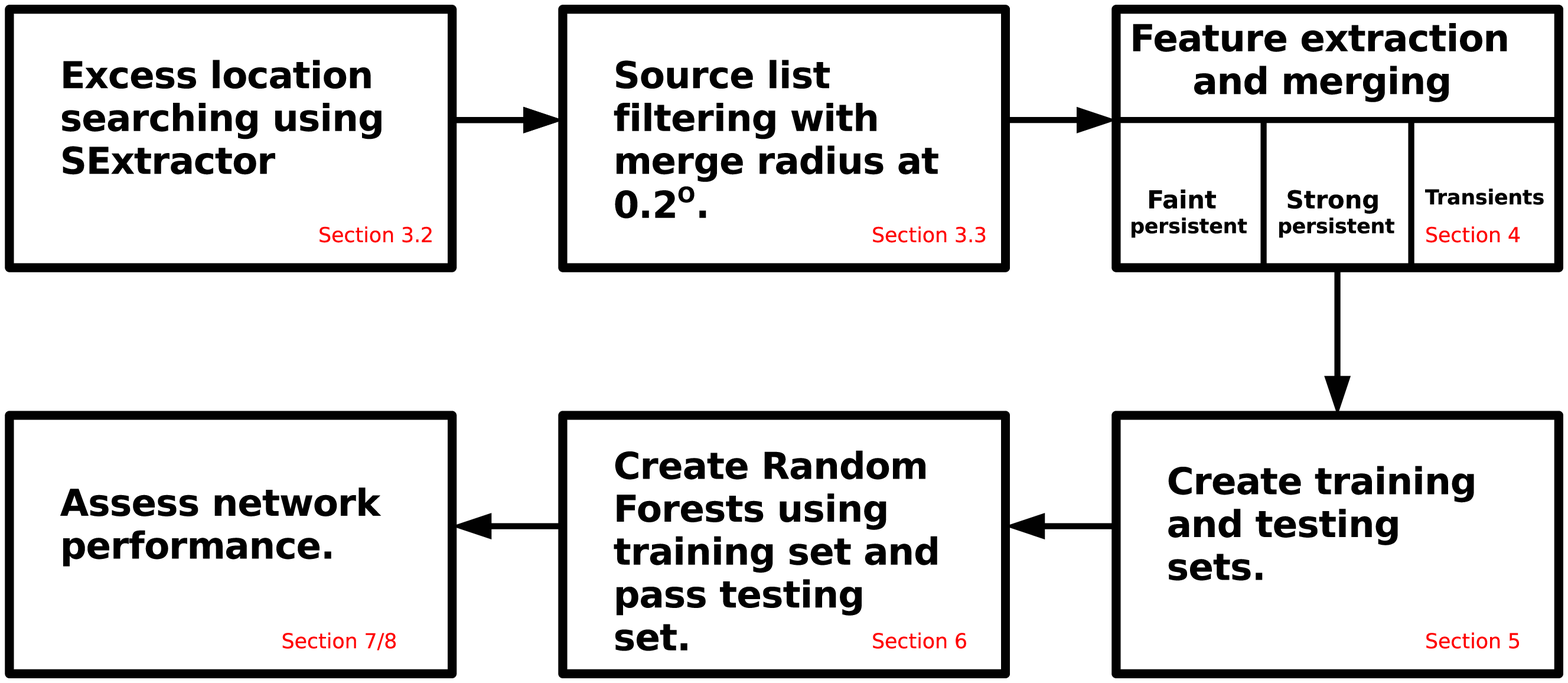}	
\caption{Flow diagram showing the levels involved in the classification procedure. The sections in which the methods are described are indicated in red.}
\label{fig:section_flow}
\end{figure*}

In Fig.\ref{fig:section_flow} we illustrate how the paper is divided into sections following the procedural flow of the algorithm. In section 2 we will give an overview of the dataset used in this work, and in section 3 we describe the methods employed in order to obtain an initial candidate list to be used by the classification network. Section 4 will describe the process of extracting features for the candidates from the IBIS/ISGRI images and the different methods to merge the features for them to be useful in a classification network. Section 5 will deal with constructing a reliable training set for our classification network to train on and a corresponding test set in order to asses the recognition performance of the network. We will describe the main Random Forest algorithm in section 6 in the context of our training and testing sets. The assessment and discussion, together with some examples of how the algorithm performs on specific objects is presented in sections 7, 8 and 9.

\section{The Dataset}
The data are collected with the low energy array ISGRI (INTEGRAL Soft Gamma-Ray Imager, \citealt{lebrun03}), consisting of a pixelated $128 \times 128$ CdTe solid-state detector that views the sky through a coded mask. The instrumental details and sensitivity can be found in \cite{lebrun03} and \cite{ubertini03}. IBIS/ISGRI generates images of the sky with $12'$ (FWHM) resolution and $\approx 3'$ source location accuracy over a $\approx 19^o$ fully coded field of view in the energy range $15-1000$ keV.

The dataset used in this work is the same as the one used in the production of the third IBIS/ISGRI soft gamma-ray survey catalogue (\citealt{bird07}), which uses  image data for the first 3.5 years of IBIS/ISGRI Core Program and public observations. The dataset used here ensures that $>70\%$ of the sky is observed with at least 10 ks exposure. The aim of this algorithm will be to ease the production of future catalogues and provide a more prompt release of information to the scientific community. More importantly the method will create less subjective catalogues.

\section{Data preparation and feature extraction}
When the input dataset to a classification algorithm is too large and/or suspected to be significantly redundant, as is the the case for the IBIS/ISGRI images, then the input data will be transformed into a reduced representation set of features (also referred to as a feature vector). As a trivial example one feature could be the significance value for a particular candidate on a particular timescale. This process is called feature extraction, or more generally dimensionality reduction. 

Feature extraction and parameter selection are the most important steps in building a reliable classification algorithm. By feature extraction we mean producing a set of variables, extracted from the IBIS/ISGRI sky images, whilst by parameter selection we mean combining these variables in order to best represent the objects we are trying to classify.  Even the most perfect classification network will not perform well if the wrong parameters are passed to it. This is why in a general scenario one has to answer the question ``what are we trying to classify?'' in order to decide what features best describe the given classes. In our case we are trying to discriminate between real sources and fake candidates within the IBIS/ISGRI images. Our features need to provide the maximum possible discrimination between real and fake sources. A feature that describes a real source is of no use if it describes a fake one in the same way. In fact features that only apply to fake sources are equally useful. Moreover they also have to take into account the nature of the artifacts caused by the imaging system, in our case the ISGRI layer on IBIS and coded masks together with the temporal nature of the gamma-ray sky.
In this section we will explore the methods employed in order to extract reliable features to be passed to the network(s) for classification. We will begin from the lowest level of the IBIS/ISGRI data, create images and mosaics, extract candidate positions, and finally extract the relevant features which will then be merged and passed into the classification network(s).

\subsection{IBIS/ISGRI data set and pipeline reduction}
The methods used for the production of the mosaic maps and science window\footnote{Each INTEGRAL pointing, of typically 2000s, is referred to as a science window} (ScW) selection are the same as those used in the third IBIS/ISGRI survey catalogue (\citealt{bird07}). In general terms, the input data set consists of all pointing data available at the end of May 2006, from revolutions (orbits) 46 to 429 inclusive\footnote{This excludes all calibration data performed in the first 45 revolutions and performance verification (PV)}. This results in more than 40 Ms of exposure time. Pipeline processing is carried out using the standard INTEGRAL analysis software (OSA 5.1; \citealt{goldwurm03}) up to the production of sky images for individual ScWs. Four primary search bands (17-30keV, 18-60keV, 20-40keV and 20-100keV) were used to optimise sensitivity and provide compatibility with previous data sets. The ScW selection for the creation of the mosaics has been done in the same way as the third IBIS/ISGRI survey catalogue, and mosaics were created for each energy band in four projections\footnote{The four projections are centred on the galactic centre, galactic anti-centre, north and south polar} and on various timescales (revolution, revolution sequence and all data).

\subsection{Excess location searching}
In order to achieve an initial candidate list we have run the source searching algorithm SExtractor (\citealt{sexref}) on the ISGRI final mosaics, revolution and revolution sequence mosaics in four energy bands. All excesses above $4.5\sigma$ in any map were extracted as possible candidates. This threshold might be too optimistic given the level of systematic noise in the maps; however, we will show how this is not a problem as the network will be able to learn and discriminate the fake candidates from the real ones. On the contrary, the threshold is too conservative for some maps where systematic background noise is very low, however at this stage it is best to have more fake candidates at the cost of recovering most of the real ones. We note that this was the global threshold employed in \cite{bird07}.

The source position measured by SExtractor relies on calculating first order moments of the source profile (referred to by SExtractor as the barycenter method). At the faintest levels source detectability will be limited to background noise, however this can be improved by applying a linear filter to the data. Moreover, in crowded regions of the sky, confusion can be avoided by applying the SExtractor Mexican hat filter. This filter convolution alters the significance of sources in the original mosaics by increasing it, deblending two (or more) close candidates. The drawback of this filter is that it sometimes creates extra ring-like candidates around apparent or real excesses, which will be extracted as a possible candidate by SExtractor. 

\subsection{Source list filtering and merging}

An initial list of 58603 excesses was extracted as described above. We need to employ some sort of filter in order to discriminate against duplicates and to remove the most obvious fake excesses. We do this by merging excesses from multiple maps by assuming two or more sources within $0.2^o$ (the IBIS/ISGRI angular resolution) from each other were actually the same, beginning from the highest significant excess. The $0.2^o$ merge radius might seem too large, however this has been chosen as a trade off between keeping the number of false positives caused by instrumental artefacts low, while still retaining the majority of objects in catalogue 3. By decreasing the merge radius we allow for more fake excess caused by the imaging system. For example bright sources in the IBIS maps tend to have propellor-like and/or ring-like structures around them sometimes extending $0.5^o$ from the source centre, and these are extracted with SExtractor. By decreasing the merge radius we allow for these to be treated as independent candidates, however by increasing the radius we allow for the candidates to be merged with the bright source from which they were created in the first place. 

\begin{figure}
\centering
\includegraphics[width=0.5\textwidth]{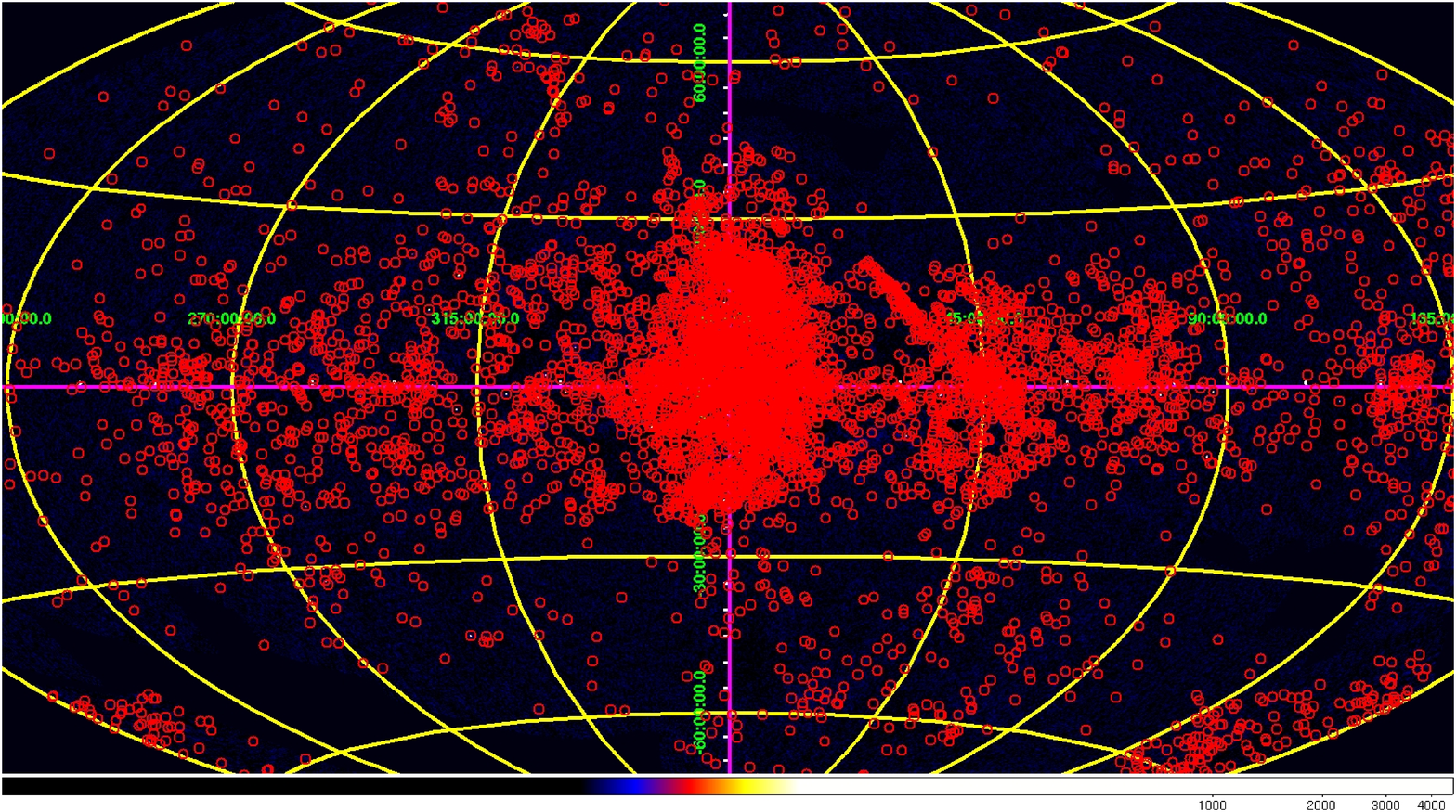}	
\caption{Galactic centre in the 18-60 keV band as seen by IBIS/ISGRI with a subset of the 7221 candidates overlayed. Candidates tend to follow the artefacts caused by the detector system. Moreover most candidates have been detected using SExtractor to a deliberately very low detection threshold, and most will turn out to be fake. The network will learn to discriminate the false excesses from the real ones.}
\label{fig:all_cands}
\end{figure}

We also eliminated all excesses that appeared only in one mosaic. This additional criterion was introduced in order to minimise the number of false positives in the final candidate list and was also the basis of the creation of catalogue 3, thus no real sources are missed by employing this cut. The final coordinates of the candidates are then taken from the highest significance excess. Thus, the initial excess list reduces to 7221 candidates, which are shown in Fig.\ref{fig:all_cands} for the galactic centre.

Out of the 421 sources identified by \cite{bird07} only 13 were not recovered with these filters. Of these, 5 were observed before revolution 46, and therefore are not present in our initial excess list. The remaining 8 were excluded due to the $0.2^o$ merge radius and will reside very close to a real source. It is possible that human intervention could recover them in the final inspection phase, however we note that these are non-trivial to recover, and new methods are being investigated in order to localise them for the production of future catalogues.

\subsection{Feature extraction}
In the context of IBIS/ISGRI source identification we have decided to use the following features from past experience in manually creating survey catalogues. First, a 2D gaussian is fitted to all ScWs where candidates might be present. We allow the gaussian to be fitted in a $9 \times 9$ pixel ($40' \times 40'$) window centered at the candidate's coordinate. The following features are then extracted:

\begin{itemize}
\item[1] Distance between gaussian center and original candidate coordinate. 
\item[2] Fitted gaussian peak (amplitude)
\item[3] Local standard deviation \footnote{Sigma clipping is employed locally to remove bright sources/structures before the calculation of the RMS}
\item[4] FWHM difference in two perpendicular directions
\item[5] FMHM ratio
\item[6] Significance value at candidate position
\item[7] Intensity value at candidate position
\item[8] Variance value at candidate position
\item[9] Residual value at candidate position
\end{itemize}

Features 1 to 5 are extracted from both intensity and significance images in all four energy ranges. A conservative cut is employed by ignoring all extracted features where the centre of the fitted gaussian is offset by more than $2.5$ pixels ($30'$) from the original candidate coordinate. In these cases the candidate is likely to be not observable in the ScW and the gaussians were fitted to background structure within the candidate region. 
In addition to the above we also extract all 9 features from the final significance mosaic maps as these will prove useful in identifying the faint persistent population. Obviously parameters such as the FWHM will depend on the kind of projection (galactic centre, galactic anti-centre, north and south polar) from which the feature is extracted. This is not appropriate as the network will then be discriminating projections rather than real FWHM. In order to deal with the problem we extract the features from the projection which has its centre closest to the candidate position, optimally minimising the distortions caused by the projections. 

On average, with large scatter, each candidate has a total of $\approx 600$ ScW pointings used in the extraction process, yielding more than 10000 features. It is clear that for many objects, in particular transients, most of the $\approx {10000}$ features will be redundant and not useful suggesting that a further step has to be employed in order to further reduce the dimensionality of our dataset. The next subsection will deal with this process called feature merging.

Once a set of relevant and reliable parameters have been chosen the problem becomes one of pattern recognition. Essentially one has created a multidimensional parameter space. In this parameter space there will be some variables which will have a greater discriminatory power than others, whilst on the other hand some combinations of two or more would be more efficient. The problem is that we are not sure which (if any) of the features are best for class discrimination and this is why one employs classification networks for pattern recognition.
   
\subsection{Relevant timescales and feature merging}
In order to reliably train a classification network, the nature and behavior of the objects one is trying to classify needs also to be taken into account. In the case of the gamma-ray sky this behavior is very diverse, and one has to define coherent subclasses that any classification network can deal with separately. After all a network which is very well trained at recognising the Crab, a bright, constant-flux source, would not necessarily perform well at recognising a faint AGN. The most obvious separation is that of faint persistent vs. strong persistent. By strong persistent we mean any objects which would be observable in one ScW pointing. On the other hand a faint persistent source might not be observable in one ScW pointing; however, its signal will still be present, and will show up in the final mosaic, for example, after having increased the exposure time on that part of the sky. To be more precise for the IBIS/ISGRI detector, a source will be observable in one ScW pointing if its flux is greater than $\approx 10$ milliCrab with a $\approx 1000$ seconds exposure. Everything with a  lower flux will need longer exposures to be observable, even though its signal will still be present in any one pointing. This is the case for most AGNs and CVs.

Another source behavior that must be taken into account when dealing with the gamma ray sky is that of transients. These objects are usually X-ray Binaries (XBs) but include a diverse set of objects as well (gamma-ray burst, supernovae). These will vary on a huge range of timescales, from being observable in only one ScW to being observable only by mosaicking several orbits of data. As one might expect these are tricky to detect as it is not known in advance what sort of timescale to expect from these objects and in particular when, in a series of pointings, to extract features from.

From here on we will refer to the definitions just described when referring to our three different source behavior types: faint persistent, strong persistent and transients. Each one of these subclasses needs to be treated independently when training as the timescales and features of each subclass vary enormously. We therefore have to tell the network what features are relevant for classification of a given subclass of sources. The danger of this approach is that we train for specific characteristics, and the detection of new source types may be inhibited. Balancing this, our subclasses are as generic as possible, which reduces the risk with specific subclasses.

In the next two subsections we will explain how the extracted features are merged in order to produce a set of merged features per network together with their respective training sets. 

\subsubsection{Faint persistent sets}
In order to deal with the faint persistent population we decide to merge the candidate features (section 3.4) by simply taking the average of, or combinations of features(see Table \ref{table_feet}). After all from our definition of faint persistent, all ScW pointings should have a signal, even if a small one. It might occur that the level of noise in any particular ScW will be much higher than the signal. As described in section 3.4, features get discarded if the gaussian fit is offset by more than $2.5$ pixels, suggesting we are looking at a ``bright'' noise structure. In our approach 12 features are used which were extracted from a ScW level and averaged as described above. We also included 6 features extracted from the final mosaic level. For a list of used features refer to Table \ref{table_feet}. It should be noted that these features have been chosen to try and mimic what an expert astronomer would consider when assessing source credibility. For example, in assessing the gaussian fit we would look at the difference between the fitted gaussian peak and the respective pixel value for a candidate source. For good fits we would expect the value to be very close to zero, whilst the value will be high for bad fits. Also note the energy bands used. For the faint persistent class we decided to use 3 energy ranges: 20-40 keV, 20-100 keV and 18-60 keV. This is because most faint persistent objects are AGN and appear in these bands from experience in compiling previous catalogues. This might inhibit the correct identification of cataclysmic variables (CVs), another subclass considered to be faint persistent but spectrally different. However INTEGRAL has not yet detected enough of these systems for them to be treated independently within the context of a source identification network. 

So in summary for each candidate in the faint persistent network we will have $12 + 6$ features merged from each used energy band, giving a total of 54 features.

\subsubsection{Strong persistent sets}
The second subclass is that of strong persistent objects. This subclass has to be treated separately from the previous, as training a network on strong persistent objects will not necessarily recover faint objects (and vice-versa). The features used for this subclass are the same as for the faint persistent subclass with the only addition of features from the 17-30keV band. This is because we think that strong persistent sources, mainly populated by XBs, are detectable through a wider spectral range. Moreover XBs are much brighter and will be detected in more energy bands. However we realise that both are persistent and that is why we essentially use the same feature timescales for both.

\subsubsection{Transient set}
\begin{figure}
\centering
\includegraphics[width=0.5\textwidth]{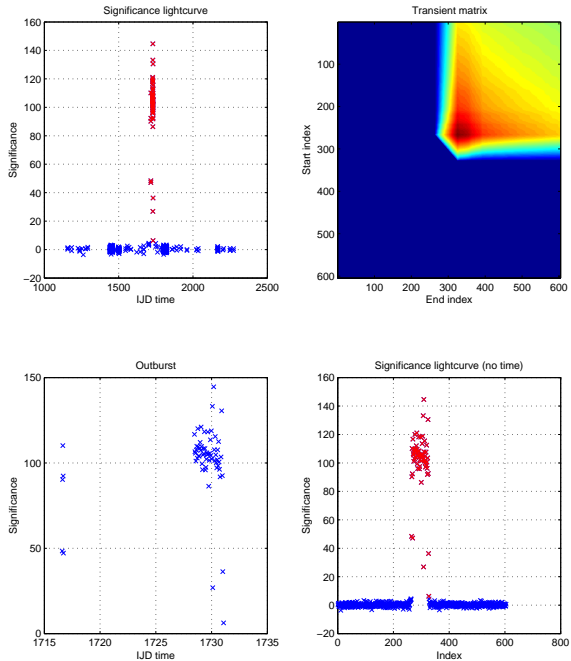}	
\caption{Graphical view of the transient matrix method applied for GT 0236$+$610. Top left: significance light curve. In red is the extracted burst. Bottom left: Significance outburst. Top right: Transient matrix. Note the indices of the maximum value. Bottom right: Significance light curve for comparison with the transient matrix (no time information).}
\label{fig:tranmat}
\end{figure}

The final subclass, transients, is the least trivial to train for, as the features are harder to define and show most variations from source to source. For this task we introduce what we call a ``transient matrix'' (TM) for the selection of ScW pointings to use.
Essentially the aim of this technique is to locate a timescale which maximises the significance of detection for transient candidates. This is important for feature merging as it will give us the features we need to average (rather than averaging all features as in the previous networks). 
Suppose the intensity light curve $I$ of a particular candidate contains $N$ points. Moreover assume each point $I(i)$ in our lightcurve has a variance $V(i)$ associated with it. We can define weights for each point in the lightcurve $w(i)=1/V(i)$. We will then create an upper diagonal $N \times N$ matrix $T$. For each row $i$ in $T$ we compute:
\begin{equation}
T(i,j)=\frac{\sum_{k=i}^j (I(k) w(k))} { \sqrt {\sum_{k=i}^j w(k)}}, \forall j\geq i
\label{tranmat_equ}
\end{equation}
where $j$ denotes the column value.
The best significance timescale is then identified by locating the row $r$ and column $c$ with the maximum value in matrix $T$ as shown in Fig.\ref{fig:tranmat} for GT 0236$+$610. This translates to a subset of the lightcurve $I$ beginning at $I(r)$ and ending at $I(c)$. Having located the beginning and end of the brightest burst/excess, we can take the mean of the features in a similar way as for the other subclasses, however this time only average those in the interval between pointings $r$ and $c$. In addition to the already defined 12 features we decide to add, for this particular network, the value $T(r,c)$. This will be an indicator of the maximum significance achievable from the light curve. 

By definition the ``transient matrix'' method will always locate a ``burst'' even if one is not there, even for faint persistent sources with no outburst. The method is meant to maximise significance, and as a result it will select all of the light curve for faint persistent sources and usually only select a small fraction of the lightcurve for fake excesses. For this reason one might think the method is biased, however we note that this method is only employed to create an additional timescale on which to merge the features; the classification of the excess will happen later in the network, which will discriminate between the real and fake excesses. The length of the outburst is not used as a feature and the coordinates $i$ and $j$ are not linear in time.

We note at this stage that all the methods described above can be performed in a totally autonomous manner without any need of human intervention.

\begin{table}
\caption{Summary of the features used within the three networks as described in section 3. In each column, a tick represents the feature being used for that particular network and viceversa for crosses.}
\begin{tabular}{c c c c}
\hline
Description & Faint &Strong &Transient \\[0.5ex]
\hline
\textbf{Significance ScW features} & & & \\

Fitted Gaussian peak & $\surd$ & $\surd$ & $\surd$ \\
FWHM difference & $\surd$ & $\surd$ & $\surd$ \\
FWHM ratio & $\surd$ & $\surd$ & $\surd$ \\\\
${\mbox{Significance}}\over\mbox{local background}$ & $\surd$ & $\surd$ & $\surd$ \\ \\
abs(${\mbox{Fitted Gaussian peak}}\over{\mbox{Significance}}$) & $\surd$ & $\surd$ & $\surd$ \\

\\
\textbf{Intensity ScW features} & & & \\

Fitted Gaussian peak & $\surd$ & $\surd$ & $\surd$ \\
FWHM difference & $\surd$ & $\surd$ & $\surd$ \\
FWHM ratio & $\surd$ & $\surd$ & $\surd$ \\\\
${\mbox{Intensity}}\over\mbox{local background}$& $\surd$ & $\surd$ & $\surd$ \\ \\
abs(${\mbox{Fitted Gaussian peak}}\over{\mbox{Intensity}}$)& $\surd$ & $\surd$ & $\surd$ \\

\\
\textbf{General ScW features} & & & \\

Variance & $\surd$ & $\surd$ & $\surd$ \\
Residual & $\surd$ & $\surd$ & $\surd$ \\
\parbox{3.5cm}{\center Maximum Significance\\ obtained from TM} & $\times$ & $\times$ & $\surd$ \\

\
\textbf{Significance mosaic features} & & & \\

Fitted Gaussian peak & $\surd$ & $\surd$ & $\times$ \\
FWHM difference & $\surd$ & $\surd$ & $\times$ \\
FWHM ratio & $\surd$ & $\surd$ & $\times$ \\
Significance & $\surd$ & $\surd$ & $\times$ \\\\
${\mbox{Significance}}\over\mbox{local background}$& $\surd$ & $\surd$ & $\times$ \\ \\
abs(${\mbox{Fitted Gaussian peak}}\over{\mbox{Intensity}}$)& $\surd$ & $\surd$ & $\times$ \\

\hline
\end{tabular}
\label{table_feet}
\end{table}

\section{Training and testing sets}
Another important issue in building a reliable classification network is the choice of reliable training and testing sets. One has to make sure that neither of these are biased towards a particular type of subclass, for example, lots of faint AGNs or lots of bright XBs or even worse not having any transients. This is one of the main reasons why we produce a training set and a corresponding classification network for each subclass. In this section we describe the methods employed to achieve this. Recall that after candidate filtering we end up with $7221$ candidates of which $408$ are present in catalogue 3.

We now have to split our candidate list into training and testing set. We have two reasonable options for producing unbiased sets:
\begin{itemize}
\item[(a)] Use the published $2^{nd}$ IBIS/ISGRI survey catalogue objects, \cite{bird2nd}, with $209$ sources for our training together with an extra $\approx 200$ fake candidates and then evaluate how the network performs in recovering the published catalogue 3 objects.
\item[(b)] Split the sky into two halves in galactic coordinates and use one half for training and the other for testing. In this case we would use catalogue 3 as a reference for what is a real source.
\end{itemize}

We decided to use option (b) as this will include some faint persistent objects only detected for catalogue 3 due to the longer exposure times and greater sky coverage compared with catalogue 2. Moreover the source types between the two catalogues are not all the same.  Option (a) will be heavily biased towards detecting more luminous sources and this will bias the network too. Moreover, by employing option (b) we can train the network for the future creation of catalogue 4. More explicitly we would expect in future to use cat$_{n-1}$ in the training for cat$_{n}$, however we require a testing set to assess the network performance and only option (b) allowed for this in an unbiased fashion.  From our initial $7221$ candidate list we now have $220$ real sources and $3114$ fake candidates for our training from the western half of the galactic sky ($0^o<l<180^o$) together with $188$ real sources and $3699$ fake candidates for testing the other half of the sky. We have chosen to split the galactic sky into West/East rather than North/South halves due to a greater similarity in the exposure times in the former case. One note of caution still needs to be addressed: some unknown fraction of what a human astronomer would consider real might actually turn out to be fake with future catalogue releases (and viceversa). This misclassification will affect the training set and therefore affect the final classification on the testing set too. Unfortunately one cannot know in advance what is real and what is not, therefore the only way to deal with this problem is having multiple iterations of the network to try and reduce the number of these false positives and  true negatives. This extra iteration step is not performed in this work but will be considered for future catalogue releases.

\section{Building the Random Forests}
As mentioned before, when building a classification network one has to take into account the nature of subclasses present in our general sets. For example one would have very limited success in correctly identifying transients if the training set only consists of faint AGN and viceversa. We have therefore decided to build three independent random forests with the three sets of features described in section 2. One will be trained only on AGN using the faint persistent set of features in order to recover faint persistent objects. A second set of forests will be trained only on XBs using the strong persistent set of features to recover bright objects, and a third set of forests, only trained on transients, using features selected by the transient matrix method. This should then allow us to recover in one or more forests all other types of sources that do not necessarily fall into the AGN/XB/transient subclasses. We reiterate that the algorithm is meant for source identification only, but as shown later, will turn out useful in source classification as well.

\subsection{How to build a Random Forest}
Classification tree methods are a good choice when the data mining task is classification or prediction. The goal of any single tree is to generate discriminatory rules that can be easily understood. Trees are constructed through a process known as binary recursive partitioning, an iterative process of splitting the data into partitions, and then splitting it up further on each of the branches (\citealt{classtrees}). We employ various classification trees in what are called Random forests. These were devised by \cite{randfo}. Essentially we build many classification trees, each tree casting a ``vote'' on a particular object.  We will build three sets of random forests using the features and training sets described previously. The final judgment as to what particular class the object is in will then be decided by the number of votes it received in each of the three forests. We reiterate that our goal is not actually to classify source types, but to maximise the efficiency for real/fake decision making.

If our training set consists of $M$ input variables (features), we will randomly choose for each tree a value $m<<M$ of variables such that each tree will be grown using only those $m$ variables. The value of $m$ is held constant for all trees grown for each subclass and is one of only two variable parameters in the network. It is responsible for two things. Increasing $m$ increases the correlations between any two trees in the subclass forest, thus decreasing its recognition strength. On the other hand increasing $m$ increases the strength of any one individual tree. A tree with a low error rate is a strong classifier, however increasing the strength of the individual trees increases the forest error rate. Reducing $m$ reduces both the correlation and the strength. Somewhere in between is an ``optimal'' range of $m$ which is usually quite wide. The other variable parameter with Random Forests is the number of trees to be grown. This has to be quite large in order to be able to use all $M$ variables through bootstrapping. There is no limit on how many trees we build in the forest as the algorithm does not overfit (\citealt{randfo}).

There are several reasons why Random Forests were used for our classification purpose. When building the network one of the main concerns was with dealing with very large datasets. Even though the IBIS/ISGRI dataset is not so large, the method presented here can deal with much larger sets. Further reasons are listed below:
\begin{itemize}
\item It can handle thousands of input variables.
\item Generated forests can be saved for future use on other data.
\item These capabilities can be extended to unlabeled data, leading to unsupervised clustering, data views and outlier detection.
\item It has the potential to give estimates of what variables are important in the classification.
\end{itemize}

Once a Random Forest has been built for a particular subclass of objects we can classify the testing set by asking how many trees in the forest will ``vote'' for that particular excess. Using this voting scheme allows us to have a feel for how confident the random forest is at assessing a particular candidate (as will be shown in the results section). If any particular excess gets enough votes in any of the networks, then it will be considered as a good candidate worth inspecting.

\subsection{Training}
As mentioned in section 3, our aim is to build 3 independent random forests in order to identify each subclass of objects separately. Recall that we have $220$ real sources and $3114$ excesses available for training (one half of the galactic sky). To build each one of the three training sets we use the classification types of the real objects published in the IBIS/ISGRI $3^{rd}$ catalogue. 

In the faint persistent case we use the $73$ AGNs in the western galactic hemisphere together with $3114$ fake excesses for our training. We cannot use all the fake candidates for a single tree or else it would bias our classification. Instead, for each tree grown in the forest, we keep the same training set of $73$ AGNs for our real sources and randomly pick $73$ fake excesses from our pool of $3114$. This ensures that no individual tree is biased towards recognising too many fake excesses, while still incorporating a wide range of them. By having this ``pool'' of fake excesses to choose from, we essentially ensure no two grown trees are the same, avoiding over-fitting. As mentioned in the previous section the only variables in our random forests are the values $m$ and the number of trees. Thus for each available set of features for a particular energy band we will choose a value $m$ together with the number of trees to grow. For example in the faint persistent network we mentioned already the use of 3 energy bands and 2 sets of features per energy band (ScW average merging and final mosaic features). We have chosen the number of trees per set to be 200 in this case, yielding a random forest with $3 \times 2 \times 200 = 1200$ trees. The value of $m$ (the random subset of features used per tree) was set to $8$ for the average features and set to $3$ for the mosaic features. These values were achieved through trial and error by maximising the accuracy of the final output given by the testing set.

In the XB case we use $46$ XBs, again from the western galactic hemisphere and use the same technique as for AGNs in dealing with the fake excess training set. In this case we chose the same value for $m$ and number of trees, however for this network we decided to include one extra energy band, yielding $4 \times 2 \times 200 = 1600$ trees.

\begin{figure*}
\centering
\includegraphics[width=1\textwidth]{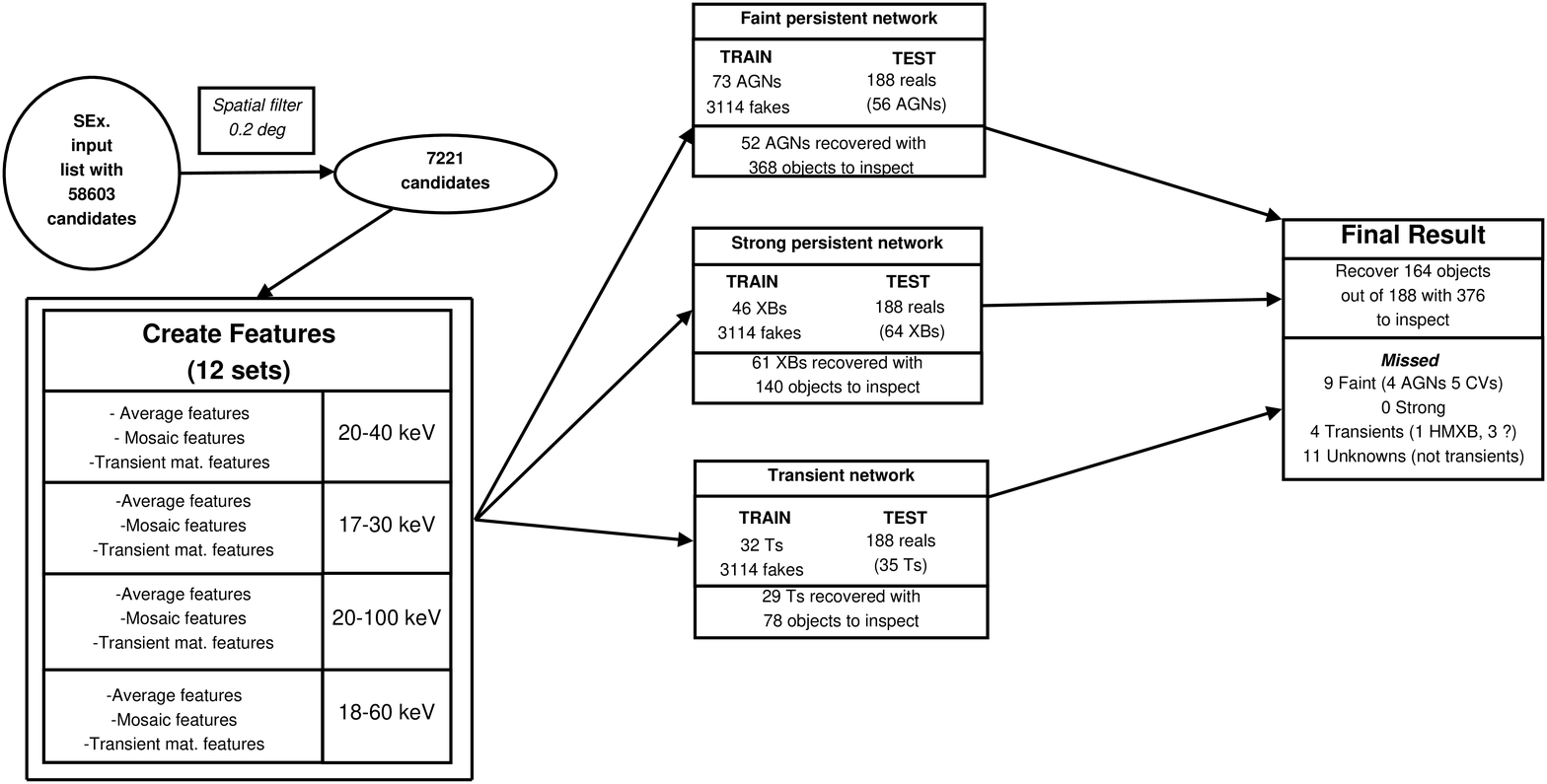}
\caption{Graphical flow diagram of the steps involved in the classification network from the extraction of the initial source candidate list to the final result. SExtractor is run on all revolutions, revolution sequences and final mosaic images. A $0.2^o$ merge radius is applied reducing the list to 7221 candidates. Features are extracted for these on a ScW level and on the final mosaic. The features are then merged in three different ways and passed to three different networks accordingly for faint and strong persistent and transients sources. In each of the network boxes we display the number of objects in the training and testing sets. In brackets we have the number of objects from the testing set in the respective subclass. Below each network box we show the result on the testing set using a $50\%$ cut on the tree votes. The final result box also applies on the testing set with a $50\%$ global cut. There we also show the missed objects and their break down.}
\label{fig:flow}
\end{figure*}

Similarly for transients we use $32$ transients for training. This network however was chosen to have a value $m=7$ and the number of trees grown per set was set to 500. This might seem very large but was used in order to have more bootstrapping from the fake candidates given the low number of transients in the training set. This yields $4 \times 1 \times 500 = 2000$ trees. We point the reader to Fig. \ref{fig:flow} for a flow diagram of the steps involved in the network and to table \ref{table_trees} for a table showing the parameters used.

\begin{table}
\caption{Summary and decomposition of the number of trees used per network.}
\begin{tabular}{c c c c}
\hline
 & Faint &Strong &Transient \\[0.5ex]
\hline
Number of trees & 200 & 200 & 500 \\
Number of energy ranges & 3 & 4 & 4 \\
Number of sets of features & 2 & 2 & 1 \\
Total number of trees & 1200 & 1600 & 2000 \\
\hline
\end{tabular}
\label{table_trees}
\end{table}

\subsection{Testing}

Recall that in our testing set we have 3887 candidates, of which 3699 are fake excesses and 188 real sources. In this section we will inspect how these candidates perform within the three independent networks. Note that all three networks had exactly the same testing set.  In order to assess the recovery performance of each of the networks we will look at how many trees voted for a particular source within the forest. If a candidate achieves $50\%$ or more of the votes then it will be considered as ``recovered''. For clarity, the analysis described here is illustrated in the flow diagram in Fig. \ref{fig:flow}, which includes the number of candidates in both training and testing sets for the three networks.

The testing set for the faint persistent network contained 56 AGNs of which 52 were recovered with the $50\%$ cut. The missing 4 AGNs were marginally below the recoverey threshold in the faint persistent network. Moreover by definition this is the network that recovers most objects. In fact a $50\%$ threshold yields 368 candidates out of the initial 3887. A lot of these will be strong persistent sources, however most will be unidentified faint persistent objects.

The strong persistent network on the other hand performed slightly better. This however is not surprising as bright sources are more easily discriminated against faint ones. Out of 64 XBs in the testing set, only three were not recovered within this network, however as we will discuss later, these get recovered in the faint persistent network. The number of candidates to inspect with the $50\%$ threshold is 140, approximately half of that produced by the faint persistent network.
  
Finally, the network producing the lowest candidate list to inspect is the transient network. This yields 78 candidates to inspect with the usual threshold. Out of 35 transients in our testing set, 6 were not recovered in this network. Of these, 2 were recovered in the faint persistent network. 

The final box in Fig. \ref{fig:flow} shows the break down of missed objects. Clearly most are unidentified, however cataclysmic variables (CVs) are also poorly recovered. As will be discussed later, this can easily be caused by not training a network for these specific source types, or they are some of the faintest and/or narrowest spectral range.

\section{Some examples}
In this section we will explore three particular sources from our testing set, each taken from a different category in our network definitions.

\begin{figure}
\centering
\includegraphics[width=0.5\textwidth]{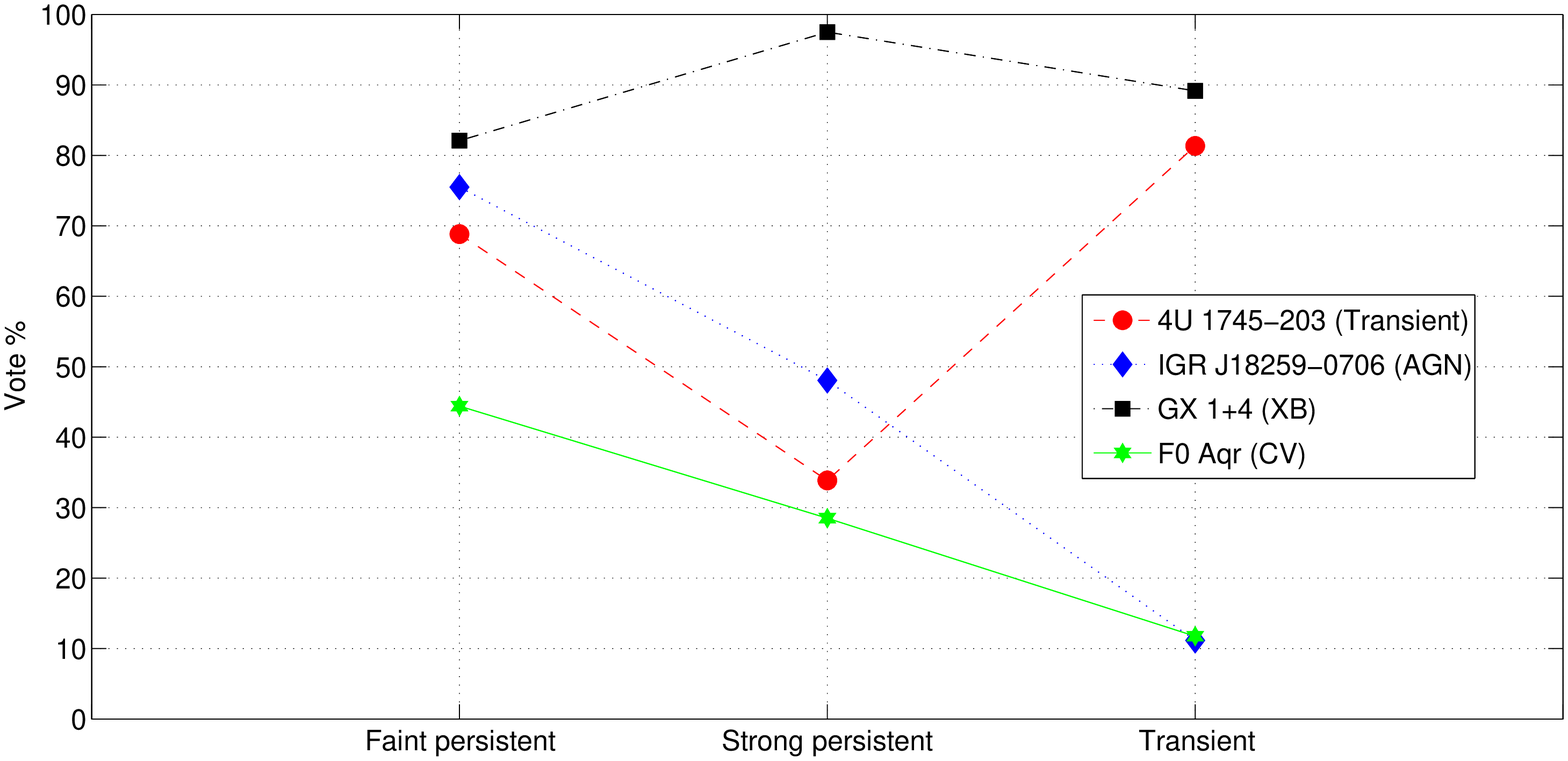}
\caption{Voting percentage obtained by three sources within the three different networks. Black dotted-dashed line: Low mass X-ray binary GX 1$+$4, Blue dotted line: AGN IGR J18259$-$0706, Red dashed line: low mass X-ray binary 4U 1745$-$203, Green solid line: Cataclysmic Variable FO Aqr}
\label{fig:exp_3}
\end{figure}

The first example is the faint persistent AGN IGR J18259$-$0706. This is a new source detected in the $3^{rd}$ IBIS/ISGRI catalogue with a maximum detection significance of $\approx 5.1\sigma$ in the 18-60 keV band and a relatively high 1570 ks exposure time. This puts it firmly in the faint persistent category. The blue curve in figure \ref{fig:exp_3} shows the percentage of votes as recorded by the three networks. It can clearly be seen that this particular example receives a greater ``confidence'' from the faint persistent network, where the curve peaks at $78\%$. This will be the source's ``global'' vote as explained in the next section. We note that at this point the network can also be interpreted as giving information about class type and not just identification. In this particular example it is clear that IGR J18259$-$0706 is classified as a faint persistent source, as the network which uses all the information available has achieved the largest number of votes: the faint persistent network.

The next example chosen is the strong persistent low mass X-ray binary GX 1$+$4 shown in black in figure \ref{fig:exp_3}. The system was detected in the $3^{rd}$ catalogue with a maximum significance of $544\sigma$ in the 18-60 keV final mosaic. Note the voting percentage difference between strong and faint persistent and transient network is very small. This is the case for most strong sources but, as observed in the previous example, not for the faint ones. In fact really strong sources tend to have high vote percentages in all three networks essentially because they are detectable on any timescale. Realistically, we only need to identify persistent vs. transient. Information on how bright they are is best obtained with other methods.

\begin{figure}
\centering
\includegraphics[width=0.5\textwidth]{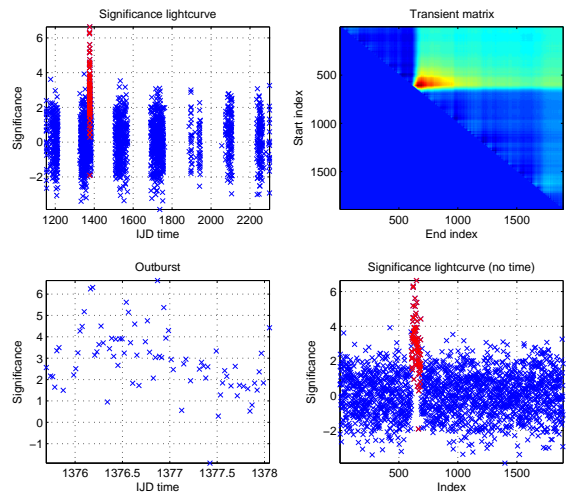}
\caption{Graphical view of the transient matrix method applied for 4U 1745$-$203. The panels are the same as those in figure \ref{fig:tranmat}.}
\label{fig:4U1745}
\end{figure}

Another example chosen is the low mass X-ray binary 4U 1745$-$203. The source was detected in the $3^{rd}$ catalogue at a significance of $20.7\sigma$ in the 20-40 keV band mosaic for revolution 120. Again, just by inspecting its corresponding red curve in figure \ref{fig:exp_3}, we can get an idea of what type of object we are dealing with had we not known in advance. The system obtains the highest score in the transient network with $84.4\%$. For this particular example we also show its transient matrix in Fig.\ref{fig:4U1745}. It can be seen that the outburst has a relatively low detection significance in any individual pointing, however, from the result of the transient matrix, the maximum significance obtained in the selected timescale is $22.4\sigma$. The source was in outburst for $\approx 3$ days, reaching a flux of $\approx 115$ mCrab. The difference between the two detection $\sigma$s is due to the fact that the transient matrix has localised as an outburst a subset of the pointings of revolution 120 instead of using them all.  

The final example chosen is the CV F0 Aqr, another weak new detection in catalogue 3. The source obtained a significance of $4.8\sigma$ in the 20-40 keV band for a 85ks exposure. This particular candidate did not achieve enough votes to be included in our ``recovered'' list, however it appears from the percentages obtained in the three networks that this is a faint persistent source. This kind of analysis can help identifying new sources even if the vote count has not passed the identification threshold.

\section{Discussion}

While section 5.3 described the performance of the network in terms of retrieval of catalogue 3 sources, here we try to quantify the network performance in more detail.

\begin{figure}
\centering
\includegraphics[width=0.5\textwidth]{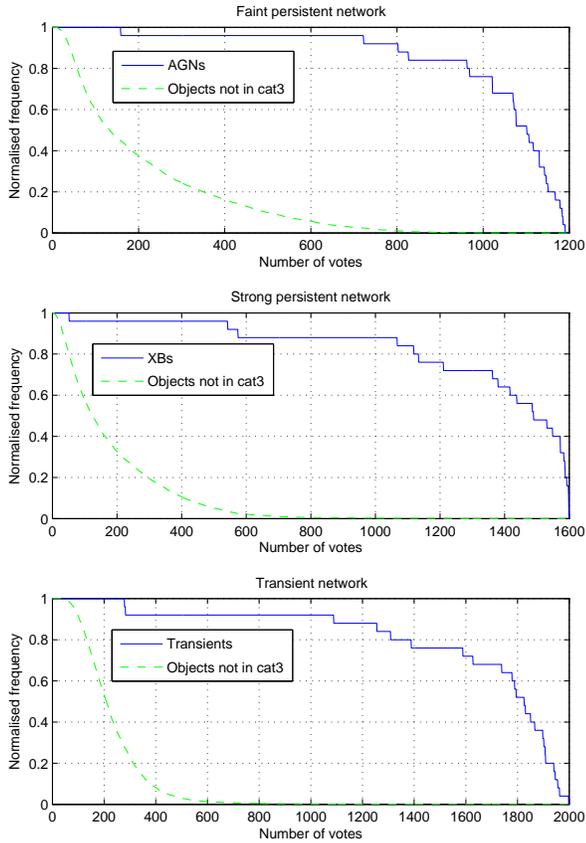}
\caption{Cumulative distribution functions for candidates within the three networks as a function of number of votes obtained. Blue solid line showing corresponding sub-class objects present in the $3^{rd}$ IBIS/ISGRI catalogue. Green dashed line are fake candidates.}
\label{fig:CDFs}
\end{figure}

Figure \ref{fig:CDFs} shows the cumulative distribution functions obtained from the three networks on the testing set. On the x-axis we have the number of votes for the network in consideration. A high vote count implies candidates are ``real'' according to the network, a low vote count rate implies the opposite. As can be seen all three networks perform relatively well in recovering their respective ``real'' objects, however contamination from the fake candidates is still present. This can be noted in the worst case scenario for the faint persistent network. This is somewhat expected as the training set for this network is by definition highly populated by low significance sources. However the transient network has performed quite well in recovering the majority of real sources whilst excluding fake ones much more easily as shown in Fig. \ref{fig:CDFs}. 

\begin{figure}
\centering
\includegraphics[width=0.5\textwidth]{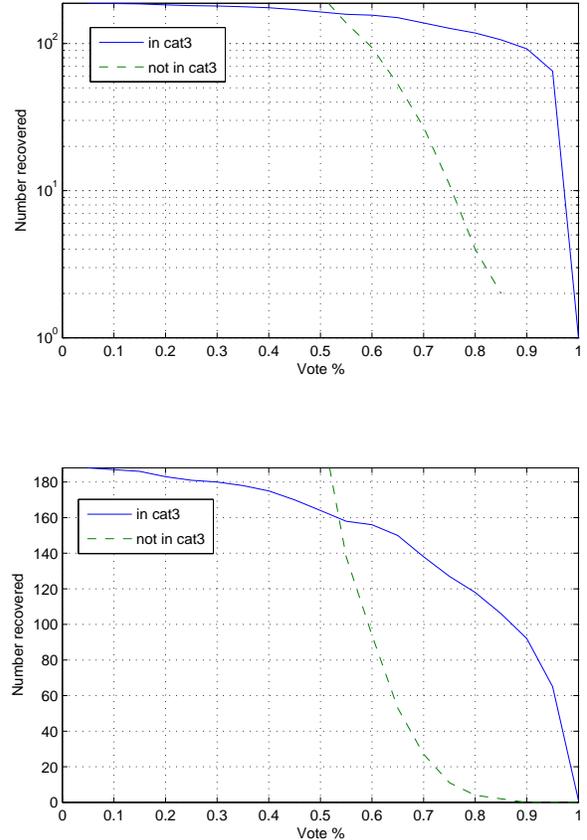}
\caption{Distribution of recovered objects present in the $3^{rd}$ IBIS/ISGRI catalogue (blue solid line) and recovered objects not present in the latest catalogue (green dashed line) as a function of global vote percentage.}
\label{fig:mainres}
\end{figure}

In order to assess the overall network performance we have to merge the results from the three networks. This is simply done by transforming the vote number for each network into percentages. Once this is done we can merge the results as a function of vote percentage as shown in figure \ref{fig:mainres}. This time the blue line represents any of the initial 188 objects found in catalogue 3 (testing set). For any candidate, the highest percentage in any of the three networks is used as a ``global'' percentage.

From visual inspection of Fig.\ref{fig:mainres} one can see that any object with $90\%$ votes or more in  will certainly be real. This includes $\approx50\%$ of the 188 real sources in the testing set. We can now query the network for the sake of reducing the amount of visual inspection for the compilation of a new catalogue. For example, if we only visually inspect and assess all candidates between $50\%$ and $90\%$ in Fig. \ref{fig:mainres}, this turns out to be 284 objects (in addition to the $92$ already accepted with $>90\%$ votes). From these we have $73$ belonging to the published catalogue 3. The remaining $23$ objects that have less than $50\%$ of the votes will be tricky to locate with this method, as below $50\%$ of the votes the number of fake candidates grows very rapidly. We note however that most of these $23$ objects are very low significance, unidentified, sources, which might even turn out to be fake excesses in future catalogue releases. On the other hand we also note that some of the fake excesses with high vote rates might turn out to be real upon further investigation.  
 
\section{Conclusion}
We have developed a reliable algorithm to aid the production of future IBIS/ISGRI gamma-ray survey catalogues. The algorithm will help produce less subjective catalogues, unbiased by human intervention. Meant for source identification, ISINA has also turned out useful in discriminating source types. We have shown how to automate the task of selecting and reducing a set of candidates from IBIS/ISGRI images.

\begin{figure}
\centering
\includegraphics[width=0.5\textwidth]{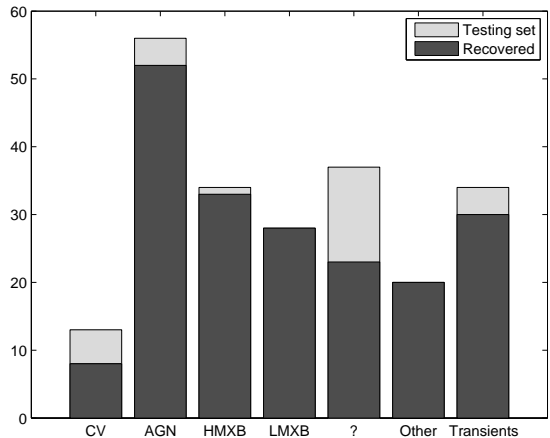}
\caption{Number of sources in the testing set classified by type (red) and recovered objects using a $50\%$ global vote cut (blue). Note that objects in the transient category are also present in there respective class types.}
\label{fig:types}
\end{figure}

The distribution of recovered objects, sorted by type, together with the objects published in the $3^{rd}$ catalogue present in the testing set are shown in Fig.\ref{fig:types}. It is clear that the majority of objects are recovered correctly with a $50\%$ global vote threshold. It is interesting to note that the only populations to suffer from a substantial decrease in recovered objects are the CVs and the unknown source types. The drop in the number of CVs can easily be explained by the fact that most of the non-recovered ones lie in crowded regions, where the systematic noise is greatest. However, we point out that missing objects in these regions does not cause a big problem for the creation of future catalogues. This is because crowded regions will be the most inspected ones, so that if the network does not recover certain objects, human intervention will. The other population to have a significant drop in the number of recovered objects is the unknown category which is a bit less trivial to assess. This is because, by definition, the only real way to determine their nature is to have longer exposure times for the regions where these are present. We also point out that both the CV population and the unknown one was not part of our training set, which might also explain the relatively low recovered rate for these. This may also have an impact in our final result as the three networks have now specialised in recovering their subclass of objects. One last observation of the general behavior of the network on the testing set is that despite the fact that the remaining classes perform well within the network, it has to be pointed out that our training and testing sets might have misclassified objects within them (false positives and true negatives). Given the nature of the classification task, the training set will always be biased towards this. However given the extremely fast data growth the problem can only get better, and these small discrepancies will systematically reduce.   

We would also like to point out the potential of such a network for exploratory data analysis in other wavelengths. The networks described here can be easily tuned to deal with different images, taken from different observatories. The features defined are quite generic, and anyway may be adjusted according to the new dataset, probably suffering from different systematic effects than the ones presented here.

Finally we would like to stress the need for such algorithms in contemporary astronomy. The network presented here has been constructed to deal with a particular dataset: namely the IBIS/ISGRI one, however multi-wavelength information might well prove useful, not only in discriminating real/fake excess but also in identifying source types. Moreover as the astronomical data size is increasing exponentially with the advent of more survey projects, larger detector areas, finer resolutions and larger fields of view the need for such algorithms is ever more pressing as human intervention for such tasks is becoming less and less feasible. 

\section*{Acknowledgements}
We acknowledge the following funding by STFC grant  PPA/S/S/2006/04459 and PP/C000714/1.

\bibliographystyle{mn2e}
\bibliography{isina}

\begin{thebibliography}{}

\bibitem[\protect\citeauthoryear{{Bertin} \& {Arnouts}}{{Bertin} \&
  {Arnouts}}{1996}]{sexref}
{Bertin} E.,  {Arnouts} S.,  1996, \aaps, 117, 393

\bibitem[\protect\citeauthoryear{{Bird}, {Barlow}, {Bassani}, {Bazzano},
  {B{\'e}langer} \& {Bodaghee}}{{Bird} et~al.}{2006}]{bird2nd}
{Bird} A.~J.,  {Barlow} E.~J.,  {Bassani} L.,  {Bazzano} A.,  {B{\'e}langer}
  G.,    {Bodaghee} A. e.~a.,  2006, \apj, 636, 765

\bibitem[\protect\citeauthoryear{{Bird}, {Malizia}, {Bazzano}, {Barlow},
  {Bassani}, {Hill}, {B{\'e}langer}, {Capitanio}, {Clark}, {Dean}, {Fiocchi},
  {G{\"o}tz}, {Lebrun}, {Molina}, {Produit}, {Renaud} \& {Sguera}}{{Bird}
  et~al.}{2007}]{bird07}
{Bird} A.~J.,  {Malizia} A.,  {Bazzano} A.,  {Barlow} E.~J.,  {Bassani} L.,
  {Hill} A.~B.,  {B{\'e}langer} G.,  {Capitanio} F.,  {Clark} D.~J.,  {Dean}
  A.~J.,  {Fiocchi} M.,  {G{\"o}tz} D.,  {Lebrun} F.,  {Molina} M.,  {Produit}
  N.,  {Renaud} M.,    {Sguera} V. e.~a.,  2007, \apjs, 170, 175

\bibitem[\protect\citeauthoryear{{Brieman}}{{Brieman}}{2001}]{randfo}
{Brieman} L.,  2001, {Machine Learning}, 45, 5

\bibitem[\protect\citeauthoryear{{Brieman}, {Friedman}, {Stone} \&
  {Olshen}}{{Brieman} et~al.}{1984}]{classtrees}
{Brieman} L.,  {Friedman} J.,  {Stone} C.,    {Olshen} R.,  1984,
  {Classification and Regression Trees}.
Wadsworth Mathematics Series

\bibitem[\protect\citeauthoryear{{Goldwurm}, {David}, {Foschini}, {Gros},
  {Laurent}, {Sauvageon}, {Bird}, {Lerusse} \& {Produit}}{{Goldwurm}
  et~al.}{2003}]{goldwurm03}
{Goldwurm} A.,  {David} P.,  {Foschini} L.,  {Gros} A.,  {Laurent} P.,
  {Sauvageon} A.,  {Bird} A.~J.,  {Lerusse} L.,    {Produit} N.,  2003, \aap,
  411, L223

\bibitem[\protect\citeauthoryear{{Lebrun}, {Leray}, {Lavocat}, {Cr{\'e}tolle},
  {Arqu{\`e}s}, {Blondel} \& {Bonnin}}{{Lebrun} et~al.}{2003}]{lebrun03}
{Lebrun} F.,  {Leray} J.~P.,  {Lavocat} P.,  {Cr{\'e}tolle} J.,  {Arqu{\`e}s}
  M.,  {Blondel} C.,    {Bonnin} C. e.~a.,  2003, \aap, 411, L141

\bibitem[\protect\citeauthoryear{{Ubertini}, {Lebrun}, {Di Cocco}, {Bazzano},
  {Bird}, {Broenstad}, {Goldwurm} \& {La Rosa}}{{Ubertini}
  et~al.}{2003}]{ubertini03}
{Ubertini} P.,  {Lebrun} F.,  {Di Cocco} G.,  {Bazzano} A.,  {Bird} A.~J.,
  {Broenstad} K.,  {Goldwurm} A.,    {La Rosa} G. e.~a.,  2003, \aap, 411, L131

\end{thebibliography}

\appendix
\label{lastpage}

\end{document}